\documentclass[%
 reprint,
 amsmath,amssymb,
 aps,
 article,
]{revtex4-2}
\usepackage{hyperref}
\usepackage{cleveref}
\usepackage{graphicx}
\usepackage{dcolumn}
\usepackage{bm}
\usepackage{svg}
\usepackage[T1]{fontenc}
\usepackage{cleveref}
\usepackage{booktabs}
\usepackage{etoolbox}
\usepackage{enumitem}
\usepackage{comment}
\usepackage{caption}
\usepackage{subcaption}
\usepackage{multirow}
\usepackage{adjustbox}

\begin{document}

\title{Tweets vs Pathogen Spread: A Case Study of COVID-19 in American States}

\author{Sara Shabani}

\affiliation{Department of Physics \& Center for Soft Matter and Biological Physics,
MC 0435, Robeson Hall, 850 West Campus Drive,
Virginia Tech, Blacksburg, Virginia 24061, USA}

\author {Sahar Jafarbegloo}
\affiliation{Department of Physics and Astronomy, University of Bologna, Bologna, Italy}

\author{Sadegh Raeisi}
\affiliation{Department of Physics, Sharif University of Technology, Tehran, Iran}

\author{Fakhteh Ghanbarnejad}
\email{fakhteh.ghanbarnejad@gmail.com}
\affiliation{School of Technology and Architecture, SRH University of Applied Sciences Heidelberg, Campus Leipzig, Prager Str. 40, Leipzig, 04317, Germany}

\date{\today}
             
\begin{abstract}
The concept of the mutual influence that awareness and disease may exert on each other has recently presented significant challenges. The actions individuals take to prevent contracting a disease and their level of awareness can profoundly affect the dynamics of its spread. Simultaneously, disease outbreaks impact how people become aware. In response, we initially propose a null model that couples two Susceptible-Infectious-Recovered (SIR) dynamics and analyze it using a mean-field approach. Subsequently, we explore the parameter space to quantify the effects of this mutual influence on various observables. Finally, based on this null model, we conduct an empirical analysis of Twitter data related to COVID-19 and confirmed cases within American states. Our findings indicate that in specific regions of the parameter space, it is possible to suppress the epidemic by increasing awareness, and we investigate phase transitions. Furthermore, our model demonstrates the ability to alter the dominant population group by adjusting parameters throughout the course of the outbreak. Additionally, using the model, we assign a set of parameters to each state, revealing that these parameters change at different pandemic peaks. Notably, a robust correlation emerges between the ranking of states' Twitter activity, as gathered from empirical data, and the immunity parameters assigned to each state using our model. This observation underscores the pivotal role of sustained awareness transitioning from the initial to the subsequent peaks in the disease progression.
\end{abstract}
\maketitle

\section{\label{sec:level1}Introduction:\protect}
The advent of new epidemic diseases prompts significant shifts in societal behavior as individuals strive to mitigate the risk of infection. The recent global challenge posed by the Coronavirus exemplifies how such events can profoundly influence various facets of life worldwide. The intricate interplay between individuals' interactions and their modified social behavior, in turn, exerts a substantial impact on the course of an epidemic~\cite{ferguson2007capturing}.

In contemporary research, the field of computational social science facilitates the meticulous documentation of behavioral responses to epidemics. For instance, a study titled "Knowledge and Determinants of Behavioral Responses to the Pandemic of COVID-19" utilized the Health Belief Model (HBM) to investigate the responses of a specific demographic, namely pharmacy students, to the Covid-19 pandemic~\cite{lv2021knowledge}. The findings underscored the variability in individuals' awareness levels regarding the disease, attributed to diverse factors. Consequently, one can posit that the level of public awareness emerges as a pivotal determinant influencing behavioral responses, and the reciprocal relationship between knowledge prevalence and disease prevalence warrants careful examination. Thus it is crucial to recognize that these processes, i.e. spreading of diseases and knowledge about it, are not isolated and impact each other. 

Interventions such as vaccination, social distancing, and improvements in healthcare accessibility can dynamically alter the spread of diseases~\cite{funk2010modelling}. These examples exemplify behavioral changes resulting from awareness about the spread of diseases. Furthermore, human behavior undergoes transformations in response to both the prevalence of the disease and prevailing societal beliefs~\cite{ferguson2007capturing}. To enhance the fidelity of epidemic models, it is imperative to incorporate human behavior as a significant factor and consider the interplay between social behavior as a response to the prevalence of disease and the subsequent changes in behavior that influence disease prevalence. A critical factor in this interplay is the concept of an \textbf{infodemic}, analogous to an information epidemic, where the dissemination of information plays a pivotal role in controlling the spread of diseases like HIV~\cite{saha2019modelling}. Governments worldwide have harnessed information dissemination efforts, especially through internet and social media platforms, to inform the public about the ongoing coronavirus pandemic~\cite{cinelli2020covid}. Focusing on accurate information dissemination, which fosters awareness and elicits behavioral responses, becomes essential in understanding and modeling the diffusion of information within society~\cite{wu2018colored, funk2010endemic,funk2009spread,wu2012impact}.

Some recent studies have explored the impact of individual behaviors, stemming from awareness, on epidemic spread within networked and non-networked models~\cite{wu2018colored,granell2014competing,wu2012impact,funk2010endemic}. However, these studies do not propose a quantitative method for studying empirical data and validating the models. To address this gap, we present a coupled Susceptible, Infectious, and Recovered (SIR) model~\cite{kermack1927contribution} that considers the interplay between the spread of diseases and the dissemination of knowledge about them. Utilizing the homogeneous mean field approximation, our approach aims not only to validate the model quantitatively but also to analyze empirical data in a rigorous manner. Moreover, our study advances beyond equilibrium-focused models~\cite{funk2010endemic,bagnoli2007risk} by scrutinizing the dynamic evolution of the system. By providing insights into changing parameters during epidemic dynamics, we offer a comprehensive understanding of the influence of awareness.

\begin{figure}[htbp]
    \includegraphics[width=0.9\columnwidth]{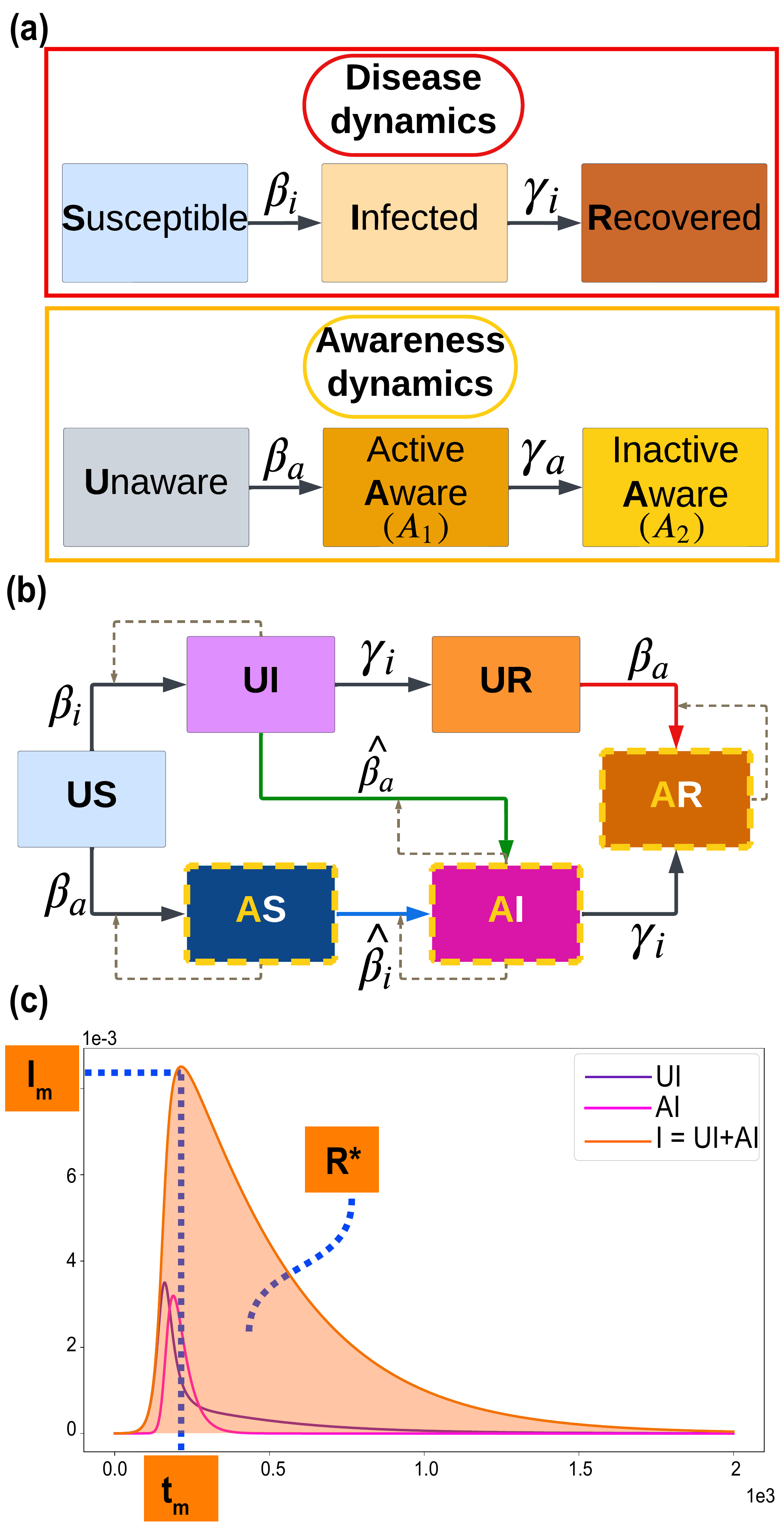}\\      
    \caption{Our model: a) A schematic of disease dynamics and Awareness dynamics. the parameters are defined in Table 1. (b) A schematic of coupled dynamics. The abbreviations are derived from the terms Aware, Unaware, Susceptible, Infected, and Recovered, indicating the state of each individual with respect to both awareness and disease status. To simplify, we consider $(A_{1}+A_{2}=A)$. The color-coded paths represent the three trajectories analyzed in the following sections to illustrate the system's progression toward equilibrium. Aware groups are also distinguished with gold color code. (c) The behavior of different Infected groups of the model, including UI, AI and I, during the dynamic. Also three important observables: R* (fraction of cumulative infected at equilibrium), $I_m$ (fraction of infected at peak time) and $t_m$ (peak time) areshown.}
    \label{fig:schematic}
\end{figure}

\section{\label{sec:level2}Model}

We develop a theoretical model as demonstrated schematically in Figure ~\ref{fig:schematic} and explain the disease and the awareness dynamics respectively, then we study how these two dynamics influence one another.\\
In order to study the disease part of our model, first we narrow our discussion to the dynamical processes in an acute infectious disease such as flu or  COVID-19 which have a short recovery period (days or weeks) followed by immunity for individuals. This process is mathematically described by the SIR model in which each agent can be Susceptible, Infected, or Recovered (see Fig.~\ref{fig:schematic}.a - Disease Dynamics). We assume the total population remains constant and use the fraction of the total population in each of the three compartments.
Additionally, we use homogeneous mean field approximation, which means we assume each agent can make contact with all other agents. That leads us to fully describe our disease dynamic by two parameters (table ~\ref{Tab:ParDes}).
\\
As represented in Fig.~\ref{fig:schematic}.a - Disease Dynamics, we have two transitions between these three groups.
\\

\begin{enumerate}[label=(\alph*)]
\item Infectious transition: Susceptible agents make contact with infected individuals and get sick at the rate of $\beta_{i}$, the infection rate.

\item Curing transition: Infected one transfer to recovered group at the rate $\gamma_{i}= \frac{1}{T_i}$, where T is the recovery period.
\end{enumerate}    

As the second component of our model, we incorporate the spread of information about the disease. As previously discussed, various dynamics can significantly influence the spread of the disease; however, our focus here is on understanding how information disseminates within society, leading to awareness about the spread of the disease. Similar to the approach in~\cite{funk2010endemic}, we model information spreading using the same framework as epidemic spreading, utilizing an SIR model. Consequently, we can investigate the parameters in the information dynamics, as outlined below and in Table~\ref{Tab:ParDes}.

\begin{enumerate}[label=(\alph*)]
\item  Awareness transition: Unaware agents get informed and move to the active aware group ($A_1$) by the transition rate ($\beta_{a}$)
\item Inactivation transition: After a specific period ($T_a$),  active aware agents move to the next phase of awareness ($A_2$) with recovery rate $\gamma_{a} = \frac{1}{T_a}$. 
\end{enumerate}

We considered a second phase of awareness, termed inactive awareness ($A_2$), wherein individuals, while remaining aware, do not actively contribute to spreading information and refrain from informing others (Fig.~\ref{fig:schematic}.a - Awareness Dynamics). This assumption is inspired by the behavior observed in the majority of people on social media platforms, as the engagement with a post would decay after a couple of days, reaching a negligible point.

\begin{table}[h!]
\centering

\begin{adjustbox}{width=0.5\textwidth}
\begin{tabular}{ |c|c|c|c|c| } 

\hline
 Notation & Description & Parameter \\
\hline
\multirow{1}{2em}{$\beta_{i}$} & transmission rate of disease for unaware group & Infection Rate\\ 

\multirow{1}{2em}{$\alpha_{1}$} & The impact of awareness on the infection & awareness Coefficient \\ 

\multirow{1}{2em}{$\hat{\beta_{i}}$} & transmission rate of disease for aware group & Effective Infection Rate\\

\multirow{1}{2em}{$\beta_{a}$} & transmission rate of awareness for susceptible group & Awareness Rate \\

\multirow{1}{2em}{$\alpha_{2}$} & The impact of infection on the awareness & Infection Coefficient\\

\multirow{1}{2em}{ $\hat{\beta_{a}}$} & transmission rate of awareness for infectious group & Effective Awareness Rate\\

\multirow{1}{2em}{$\gamma_{i}$} & recovery rate for disease & infection Recovery rate\\ 

\multirow{1}{2em}{$\gamma_{a}$} & recovery rate for awareness & Awareness Recovery Rate\\

\hline
\end{tabular}
\end{adjustbox}
\caption{Description of the Model Parameters}
\label{Tab:ParDes}
\end{table}
Finally, we expand our model to consider how these two dynamics, disease and awareness, can mutually influence each other, studying the entire scenario as coupled dynamics. As discussed in~\cite{funk2010modelling}, this influence can manifest as changes in individual states, model parameters, or the overall structure. Given that the spread of information can lead to behavioral changes, we assume these impacts are reflected in alterations to the transmission rates of both dynamics. Specifically, awareness is assumed to effectively reduce the transmission rate for epidemic spreading, inducing partial immunization. Conversely, the disease can modify the awareness rate in the agent. Thus the infection rate for the aware group and the awareness rate for the infected group change according to the following factors, respectively: $\hat{\beta_{i}} = \alpha_{1}\beta_{i}$, $\hat{\beta_{a}} = \alpha_{2}\beta_{a}$. 

The awareness alters the situation for an agent in a way that reduces the chance of getting infected. So we should set the range of awareness Coefficient, $\alpha_1$, to [0,1]. Interactively, the infection plays a crucial role in spreading the information itself, and consequently, infected agents become aware at a higher rate. Therefore, infection Coefficient, $\alpha_2$, should be greater than 1. 

The coupled dynamic of the epidemic and infodemic is illustrated in Fig.~\ref{fig:schematic}.b. In this schematic, we use abbreviated form of unaware group by 'U', while we merge the two state of awareness, active, i.e. $A_1$ and inactive aware, i.e. $A_2$, and show both groups by 'A', and disease situation follows the abbreviated form of SIR model. As shown the total population is divided by the compartments based on their disease and awareness situation. Now we can write the ODE equations for this coupled system as follows:

\begin{align*}
\frac{d[US]}{dt} = -\beta_{a}[A_{1}][US] - \beta_{i}I[US]\hspace{3.45cm}
\\
\frac{d[A_{1}S]}{dt} = \beta_{a}[A_{1}][US] - \hat{\beta_{i}}I[A_1S] - \gamma_{a}[A_{1}S]\hspace{1.8cm}
\\
\frac{d[A_{2}S]}{dt} =  \gamma_{a}[A_{1}S] - \hat{\beta_{i}}I[A_{2}S] \hspace{3.9cm}
\label{eqn:eqlabel_1}
\end{align*}

\begin{align}
\begin{split}
\frac{d[UI]}{dt} = \beta_{i}I[US] - \hat{\beta_{a}}[A_{1}][UI] - \gamma_{i}[UI] \hspace{1.99cm}
\\
\frac{d[A_{1}I]}{dt} = \hat{\beta_{a}}[A_{1}][UI] + \hat{\beta_{i}}I[A_{1}S] - \gamma_{a}[A_{1}I] - \gamma_{i}[A_{1}I] 
\\
\frac{d[A_{2}I]}{dt} = \gamma_{a}[A_{1}I] + \hat{\beta_{i}}I[A_{2}S] - \gamma_{i}[A_{2}I] \hspace{2.1cm}
\end{split}
\end{align}
\begin{align*}
\frac{d[UR]}{dt} = -\beta_{a}[A_{1}][UR] + \gamma_{i}[UI] \hspace{3.7cm}
\\
\frac{d[A_{1}R]}{dt} = \beta_{a}[A_{1}][UR] + \gamma_{i}[A_{1}I] - \gamma_{a}[A_{1}R] \hspace{2cm}
\\
\frac{d[A_{2}R]}{dt} = \gamma_{a}[A_{1}R] + \gamma_{i}[A_{2}I] \hspace{4.1cm}
\end{align*}

\vspace{0.6cm}

As the next step, we define three quantitative observables in our model and investigate how varying parameters impact them. These observables are crucial both theoretically and from a practical epidemiological standpoint.

\begin{itemize}
    \item Fraction of Total Infected: (R*)
        \newline The fraction of individuals who got infected during the pandemic until the wave of infections is over. 
        
    \item Peak Value of Daily Infected: ($I_m$)
        \newline The fraction of individuals who got infected at the peak of the daily infected 
        
    \item Peak Time of Daily Infected: ($t_m$)
        \newline The time when the peak of daily infected takes place       
\end{itemize}

The first observable, R*, is important as it informs us about the equilibrium of the epidemic dynamics — or, in epidemiological terms, herd immunity — which indicates how extensively the pandemic has spread through society by the end of the epidemic.

The second and the third observables, $I_m$ and $t_m$, also tell us the value and time of the daily infected curve's maximum. These are considerable observables since the $I_m$ can determine the sharpness of the daily infected curve in fixed R* and $t_m$ determines the time of this peak. Comparing these two fractions with the health system capacity is practical and can provide us with some information by which we can apply appropriate policy and react properly to an epidemic. These three observables are shown in Fig.~\ref{fig:schematic}.c, which illustrates the evolution of the unaware infected, the aware infected, and the whole infected group until the dynamic reaches equilibrium.

As mentioned, we study the model in its equilibrium, that is, when it reaches the absorbing state with no active infections, by analyzing R*. Also, we will investigate how the model reaches equilibrium by dividing R* into three groups, based on their awareness situation when they get infected.

\begin{figure*}
    \includegraphics[width=1.02\textwidth]{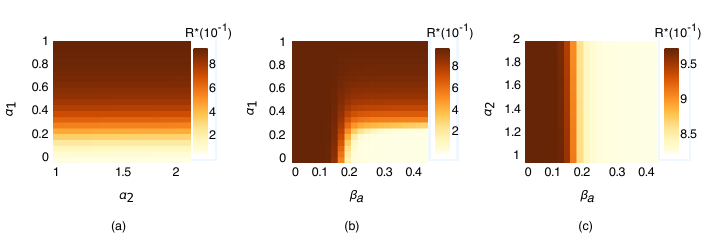}
    \caption{Fraction of Total Infected at Equilibrium(R*). Integrating equation \ref{eqn:eqlabel_1}, by initial conditions set to $[AS]=10^{-6}$, $[UI] = 10^{-6}$ and other variables set to zero. We set $\beta_i = 0.15$ and $\gamma_i= \frac{1}{14}$ and $\gamma_a= \frac{3}{2}$. Time step has been set to h=0.5 for 2000 iterations. Plot (a): $\beta_a= 0.15$ Plot (b): $\alpha_2=1$ which removes the effect of disease on awareness while $\alpha_1$ and $\beta_a$ vary. Plot (c): $\alpha_1= 0.5$ which presents half immunization. Note that the range of color bars are not the same for all panels.}
    \label{fig:Rstar}
\end{figure*}

\begin{figure*}
    \includegraphics[width=1.02\textwidth]{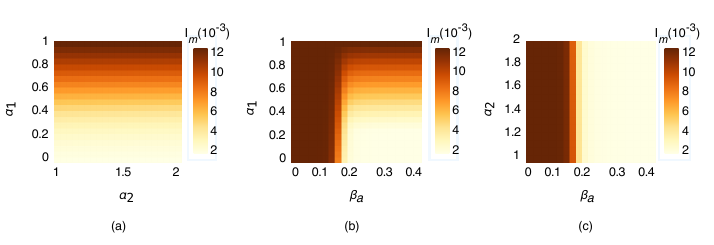}
    \caption{Peak Value of Daily Infected ($I_m$). Equations, initial conditions and the parameters are the same as the Figure \ref{fig:Rstar}}
    \label{fig:I_m}
\end{figure*}

\begin{figure*}
    \includegraphics[width=1.02\textwidth]{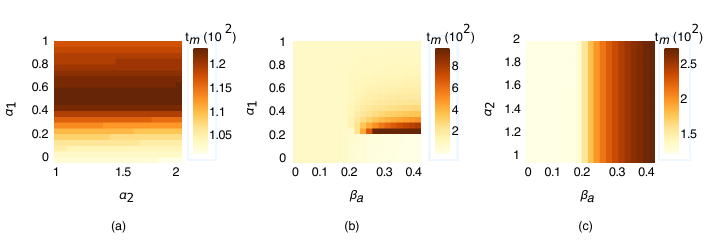}
    \caption{Peak Time of Daily Infected($t_{m}$). Equations, initial conditions and the parameters are the same as the Figure \ref{fig:Rstar}}
    \label{fig:t_m}
\end{figure*}

\section{\label{sec:citeref}Simulation Results}

Although our model is general, by constraining certain parameters, we can apply it to the analysis of a real epidemic influenced by the spread of information in society. Inspired by the recent pandemic, we fix three parameters of our model—$\beta_{i}$, $\gamma_i$, and $\gamma_a$—based on previous studies of the COVID-19 outbreak. First, we base our analysis on the determined values of two epidemic parameters, $\beta_i$ and $\gamma_i$. Previous results~\cite{linka2020reproduction} indicate that the value of $R_0$ is equal to 4. Moreover, By assuming the value of recovery period of COVID-19 is equal to 14 days (2 weeks), we fix the value of $\beta_i$, knowing that $\gamma_i = \frac{1}{14}$ and $\beta_{i} = R_0 \times \gamma_i$. Furthermore, we fix $\gamma_a$ in the information dynamic. We assume it takes approximately ten days for an aware agent to transfer into the second phase and become an inactive aware agent.
Now, by fixing these three parameters, we can numerically solve equation 1 and investigate how the behavior of the dynamic changes as we change the parameters $\alpha_{1}$, $\alpha_{2}$, and $\beta_{a}$. We set the initial conditions of the simulation into $A_1S=10^{-6}$ and $UI=10^{-6}$ in order to be a good approximation for majority of empirical data. We now examine how the three observables—$R^*$, $I_m$, and $t_m$—vary with respect to the parameters $\alpha_{1}$, $\alpha_{2}$, and $\beta_{a}$.

In the first panels (a) of Figures~\ref{fig:Rstar}, \ref{fig:I_m}, and \ref{fig:t_m}, we fix $\beta_a$ equal to $\beta_i$—meaning the information dynamics have the same transmission rate as the disease—and examine how the observables change as $\alpha_1$ and $\alpha_2$ vary.
In the second panels (b) of the same figures, we fix $\alpha_2 = 1$, which implies that the epidemic does not influence the infodemic, and vary $\alpha_1$ and $\beta_a$.
In the third panels (c), we fix $\alpha_1 = 0.5$ (representing partial immunization) and explore the effects of varying $\alpha_2$ and $\beta_a$.
In all panels of Figures~\ref{fig:Rstar}, \ref{fig:I_m}, and \ref{fig:t_m}, the values of the three observables are represented using color-coded heatmaps, with corresponding color bars provided for each panel.

\subsection{Fraction of Total Infected}
Firstly, we focus on the fraction of total infected, which is demonstrated in Fig.~\ref{fig:Rstar}. By examining Fig.~\ref{fig:Rstar}(a), we observe that R* does not vary significantly as $\alpha_2$ changes. However, decreasing $\alpha_1$ results in lower values of R*.  Namely, when $\alpha_1$ decreases, immunization becomes stronger, and the total number of infected individuals decreases. We can infer from this result that $\alpha_2$ does not effectively change the destiny of the dynamic. The influence that the disease has on the information dynamics and changes in $\hat{\beta_{a}}$ does not affect the fraction of total infected, as we demonstrate analytically in Appendix~\ref{app:R_0}. On the other hand, decreasing $\alpha_1$ and subsequently, decreasing $\hat{\beta_{i}}$ can reduce effective $R_{0}$ below 1 for the aware group and control the epidemic (Appendix~\ref{app:R_0}). 
\\
Fig.~\ref{fig:Rstar}(b) depicts that as $\beta_a$ increases and the information spread more rapidly, the fraction of total infected decreases. Also, we observe that as $\alpha_1$ decreases and the immunity gets stronger, the fraction of total infected decreases.

Furthermore, we can infer from this panel that when $\beta_a$ is greater than $\beta_i$ (0.15), information propagates faster than the disease. If $\alpha_1$ is sufficiently small (less than 0.25) (the area in the bottom right), it can reduce the effective reproduction ratio below 1 for the aware group, effectively controlling the epidemic. In other words, in this area, $\hat{\beta_{a}}$ is greater than $\hat{\beta_{i}}$. This indicates that the first term of the first equation is greater than the second term. Consequently, the transition from the [US] state to the [$A_1S$] state occurs at a higher rate than the transition from the [US] state to [UI], resulting in a lower fraction of unaware individuals who get sick compared to aware individuals. Moreover, as previously mentioned, in this region, $\alpha_1$ is below 0.25, and the reproductive rate ($R_0$) for the aware group is less than 1, indicating that the infection is not in the epidemic phase for this dominant group.
In Fig.~\ref{fig:Rstar}(c), we set $\alpha_1$ to 0.5, representing half immunization. This plot confirms that $\alpha_2$ has no significant effect on the fraction of infected individuals at equilibrium. When we observe the change in $\beta_a$ in Fig.~\ref{fig:Rstar}(c), we see that the fraction of total infected decreases as the $\beta_a$ parameter increases.
\subsection{Peak Value of Daily Infected}
We can go further and see how these parameters can be influential during the dynamic. We investigate how the peak value of daily infection ($I_{m}$) changes as the parameters change in Fig.~\ref{fig:I_m}. We see the same patterns as total infection for ($I_{m}$), shown in Fig. \ref{fig:Rstar}. So we can conclude that the behavior of dynamic in the peak and fraction of infected cases we have in the peak follow the the behavior of system at its equilibrium. In Fig.~\ref{fig:I_m}(a) and Fig.~\ref{fig:I_m}(b) We observe a smaller peak for smaller $\alpha_1$. Fig.~\ref{fig:I_m}(c) represents that as the $\beta_{a}$ increases, a smaller peak. Similarly, we see $\alpha_2$ has not influenced on $I_{m}$ and a transition occurs near $\beta_{a} = 0.15$. 

\begin{figure*}
    \centering
    \includegraphics[width=1.02\textwidth]{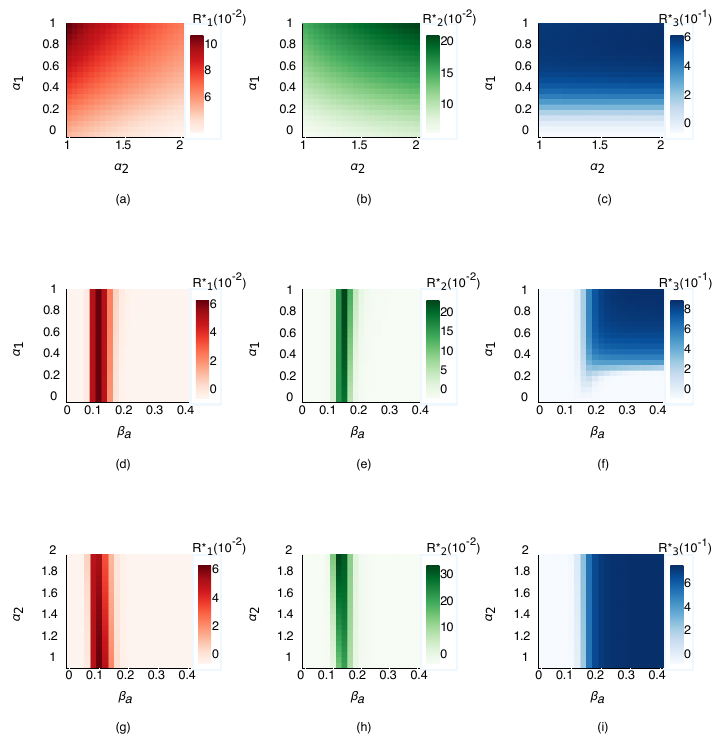}

    \caption{How to Reach Equilibrium.  
This figure illustrates the effect of key model parameters on the final fraction of individuals in three distinct transition groups. Each group is color-coded, corresponding to the colored paths highlighted in Fig.~\ref{fig:schematic}. The equations, initial conditions, and baseline parameters used here are identical to those in Fig.~\ref{fig:Rstar}. The three groups are defined as follows:  
\textbf{Red:} Individuals who become infected while unaware and only become aware after recovery ($UR \rightarrow AR$).  
\textbf{Green:} Individuals who become aware upon infection ($UI \rightarrow AI$).  
\textbf{Blue:} Individuals who are aware before becoming infected ($AS \rightarrow AI$). Note that the color bar ranges are not uniform across all plots.  Subfigures illustrate the results across three parameter sweeps:  Panels (a–c): $\beta_a= 0.15$   Panels(d–f): $\alpha_2= 1$ and Panels  (g–i): $\alpha_1= 0.5$}.
    \label{fig:Equilibrium}
\end{figure*}

\subsection{Peak Time of Daily Infected}
Finally we study the third observable, peak time of daily infected $t_{m}$ by focusing on Fig.~\ref{fig:t_m}(a).
This figure illustrates that for a constant $\beta_a$=0.15, we observe that the peak time of daily infected cases ($t_{m}$) increases as $\alpha_{1}$ decreases and immunization gets more robust from $\alpha_{1}$=1, no immunization, until the region where we have half immunization, $\alpha_{1}$=0.5. For values of $\alpha_1$ lower than 0.5, we see a gradual decrease of the time of the maximum. That is to say, we do not have a monotonic behavior, and the latest time of maximum will be reached in the half immunization region. Also, this plot represents that $\alpha_2$ would not considerably affect the time of the maximum in compere with $\alpha_2$ variation. We can conclude that for a very low immunization where $\alpha_1$ is near to one, the maximum takes place very soon an the high number of individuals get infected very fast. Also, for a strong immunity, where $\alpha_1$ is near to zero, the epidemic has died out very soon, so the peak, which has a low value, takes place at the beginning of the dynamics.  

In Fig.~\ref{fig:t_m}(b), we observe the emergence of a region in parameter space where $t_m$ increases significantly.

In Fig.~\ref{fig:t_m}(c), we can see that higher transition rates for infodemic can hinder the epidemic and delay the maximum. Because the information spreads more rapid than disease, the peak of the infection delays in the dynamic. Similarly, we observe no changes in peak time for different value of $\alpha_1$ neither.

Looking at Fig.~\ref{fig:Rstar}, \ref{fig:I_m}, and \ref{fig:t_m} at the same time, we understand how our model behaves in parameter space. We see that the fraction of total infected and the peak value of daily infected follow the same pattern, and if we want to control the pandemic, we should increase our awareness-based immunization and reduced $\alpha_1$ below 0.25 (resulting in a drop in $R_0$ for the aware group below 1.0), and $\beta_{a}$ should be higher than $\beta_{i}$. By looking at Fig.~\ref{fig:t_m}, The half immunization, or the area where $\alpha_1$ = 0.5, becomes more essential than $\alpha_1$ = 0.25 if we desire to limit the peak time of daily infection which decreases whether we weaken or boost the immunization from 0.5. Although Fig.~\ref{fig:Rstar},and Fig.~\ref{fig:I_m} demonstrate a controlled pandemic in areas where $\beta_{a} > \beta_{i}$, Fig.~\ref{fig:t_m}(b) shows that for these values of  $\beta_{a}$, the peak time of daily infected is relatively late if $\alpha_1$ = 0.25. We deduce from this discussion that the information indicated in Fig.~,\ref{fig:Rstar}, and Fig.\ref{fig:I_m} are better to concentrate on. For example, in the area where the peak time occurs relatively late, Fig.~\ref{fig:t_m}(b) and Fig.~\ref{fig:t_m}(c), the peak value is modest enough that we may argue that the late peak time is negligible. Also, if the immunization is strong, we have an early peak time, but because the peak value is small, it would not be challenging to interface.

We can define three different strategies based on the previous analysis on these three observables. If we want to follow the strategy which reduce the fraction of total infected, we should reduce the immunization parameter, $\alpha_1$, below 0.25 and accelerate the awareness spreading and increase $\beta_{a}$ up to more than $\beta_{i}$. If our strategy follow the aim of peak value reduction, we need to decrease $\alpha_1$ in to half immunization or stronger($\alpha_1$<0.5) and also $\beta_{a}$ should be higher than 
$\beta_{i}$ as previous strategy. In the third strategy, if we want to control the peak time, we should note that earliest peaks takes place for a very strong or very weak immunization($\alpha_1$ near to one or zero) and latest peak takes place for half immunization($\alpha_1$ around 0.5). Also for $\alpha_1$ around 0.25, if $\beta_{a}$ is higher than  $\beta_{i}$ we have a late peak time but low peak value. 

\subsection{Three Ways Reaching the Equilibrium}
In order to discuss the effect of the dynamic more precisely, we can divided infected people into three groups. The first group is those who get infected and remain unaware, and after the recovery, they get aware ($UR\xrightarrow[]{}AR$)($R^*_1$, shown in Fig.~\ref{fig:Equilibrium} in red). The second group refers to people who get the disease and, by that, get aware.($UI\xrightarrow[]{}AI$)($R^*_2$, shown in fig.~\ref{fig:Equilibrium} in green). The third ones are aware and then get the disease ($AS\xrightarrow[]{}AI$)($R^*_3$, shown in fig.~\ref{fig:Equilibrium} in blue). Now we want to see how the fraction of total infected for these three groups is affected by changing parameters. Look at Fig.~\ref{fig:Equilibrium}. (A more detailed explanation of each group will be given later.)

In Figs.~\ref{fig:Equilibrium}, panels a, b and c we fixed $\beta_{a} = 0.15$. In Fig.~\ref{fig:Equilibrium}(a) We can tell that for group 1, as $\alpha_2$ increases, the fraction of group 1 decreases, and the fraction of group 2 increases(Fig.~\ref{fig:Equilibrium})(b). This result was expected since, with increasing $\alpha_2$, we increased the impact of the disease on awareness, and infected people became informed at a higher rate. So, the fraction of group a, which remains unaware even after their infection, must decrease, and subsequently, the fraction of people who get informed by means of their infection increases. Fig.~\ref{fig:Equilibrium}(c) also express that changing $\alpha_2$ has no effect on group three. awareness of this group is independent of the disease so this result is expected. Also, we can see from these three plots that as $\alpha_1$ increases, the fraction of all three groups increases because the immunization of people gets weaker and as shown in Fig.~\ref{fig:Rstar} and discussed in section (A), the number of infected increases in general.

In Figs.~\ref{fig:Equilibrium}(d) and ~\ref{fig:Equilibrium}(g), we show that for $\beta_a$ approximately around 0.1, the dominant group is 1. Information spreads very slowly, and consequently, people are informed far after they have the disease. According to Figs.~\ref{fig:Equilibrium}(e) and ~\ref{fig:Equilibrium}(h), as $\beta_a$ increases, group 2 becomes prevailing, and most people get informed when they get the disease. in these four figurs, in contrast with Figs.~\ref{fig:Equilibrium}(a), ~\ref{fig:Equilibrium}(b), and ~\ref{fig:Equilibrium}(c), changes in the fraction of group one and two are not recognized by varying $\alpha_1$ and $\alpha_2$, in results of huge changes in response to $\beta_a$ alteration.

Figs.~\ref{fig:Equilibrium}(f) and ~\ref{fig:Equilibrium}(i) depict that as $\beta_a$ exceed this region and information propagates in a higher rate of the disease, the third group has become the major group since individuals get informed sooner than getting infected. 
Fig.~\ref{fig:Equilibrium}(f) represents a white region in high values of $\beta_a$ and low values of $\alpha_1$ for the third group. This tells that a very low fraction of individuals (as discussed in Fig.~\ref{fig:Rstar}) get sick due to strong immunization  and information spreads rapidly. Fig.~\ref{fig:Equilibrium}(i) confirms that $\alpha_2$ has no influence on the fraction of group three.

\subsection{Correlation}
\begin{figure}[t]
    \centering  
    \includegraphics[width=.97\linewidth]{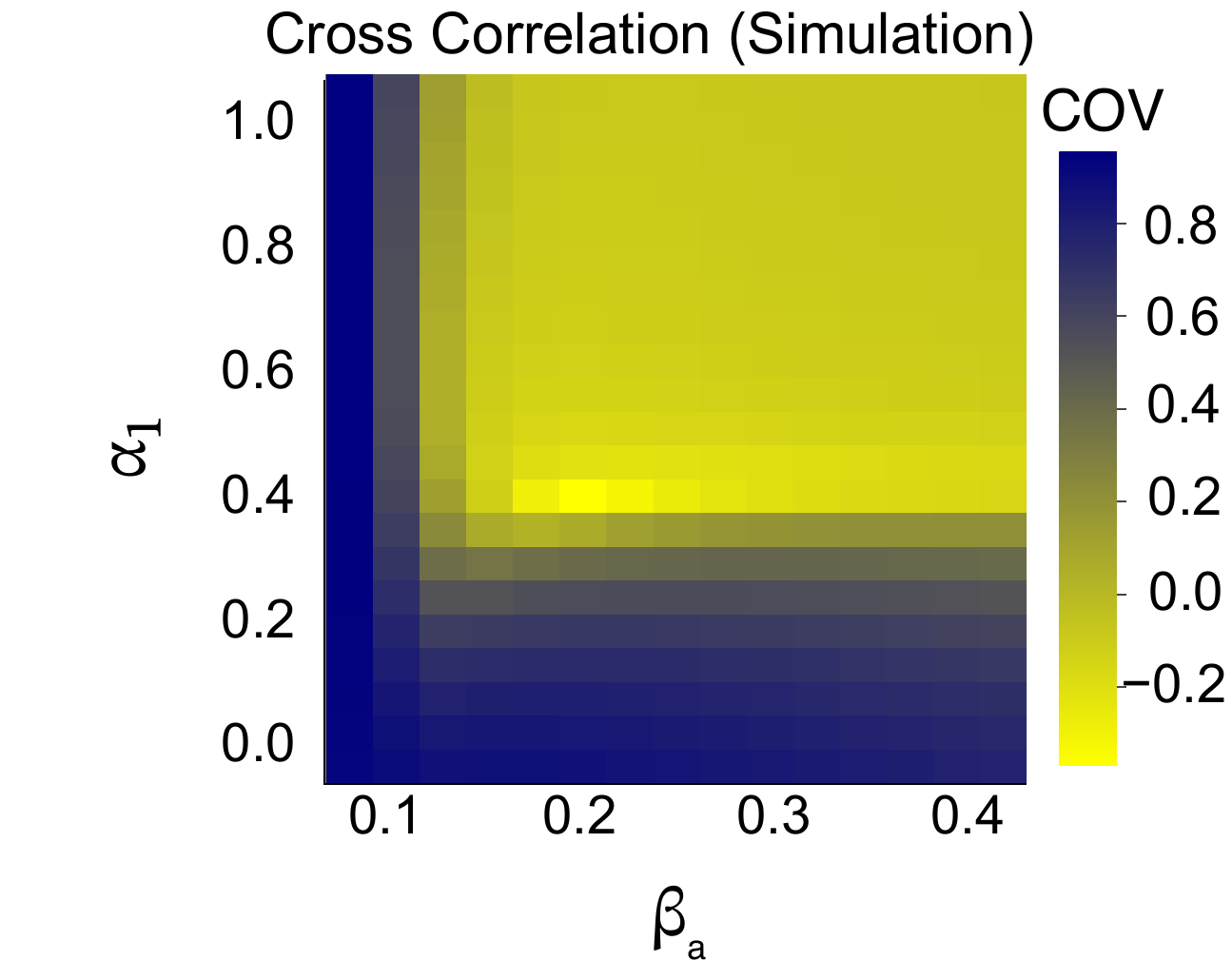}
    \caption{
    Cross Correlation. The covariance $\mathrm{COV}[A, I]$ obtained from simulations, where Eq.~ \ref{eqn:eqlabel_1} is solved for varying values of $\alpha_1$ and $\beta_a$, while keeping the other parameters fixed. Specifically, we set $\beta_i = 0.15$, $\gamma_i = 1/14$, and $\gamma_a = 3/2$. The initial conditions are set to $[AS] = 10^{-6}$ and $[UI] = 10^{-6}$.}
    \label{fig:Cross}
\end{figure}

Now we want to investigate whether the fraction of aware agents correlates with the fraction of infected agents or not. That means there is a relation between the epidemic and infodemic spreading. To do so, we apply the Pearson's correlation coefficient as follows:
\vspace*{-0.1mm}
\begin{align}
\rho_{A,I} = \frac{COV[A,I]}{\sigma_{A} \sigma_{I}}
\label{eq:conditionalentorpyresult}
\end{align}
By applying the Pearson's correlation on daily infected case and daily aware case for different coupling parameters ($\alpha_{1}$ and $\beta_{a}$), we investigate how the two dynamics are correlated for different parameters. In Fig.~\ref{fig:Cross}, we observe that smaller values of $\beta_{a}$ and $\alpha_1$ correspond to a stronger correlation. As expected, if the transmission rate of awareness is less than epidemic, we do not have a strong correlation, also for weak immunization, even for high value of $\beta_{a}$, we see week correlation as the coupling parameters is weak and near to 1. The strongest correlation can be seen in those area where either $\beta_{a} = \beta_{i}$, the two dynamics spreads in equal rates, or $\alpha_{1}$ is near to zero and have a strong immunization. 

\begin{figure}[htbp]

    \centering
    \includegraphics[width=0.9\columnwidth]{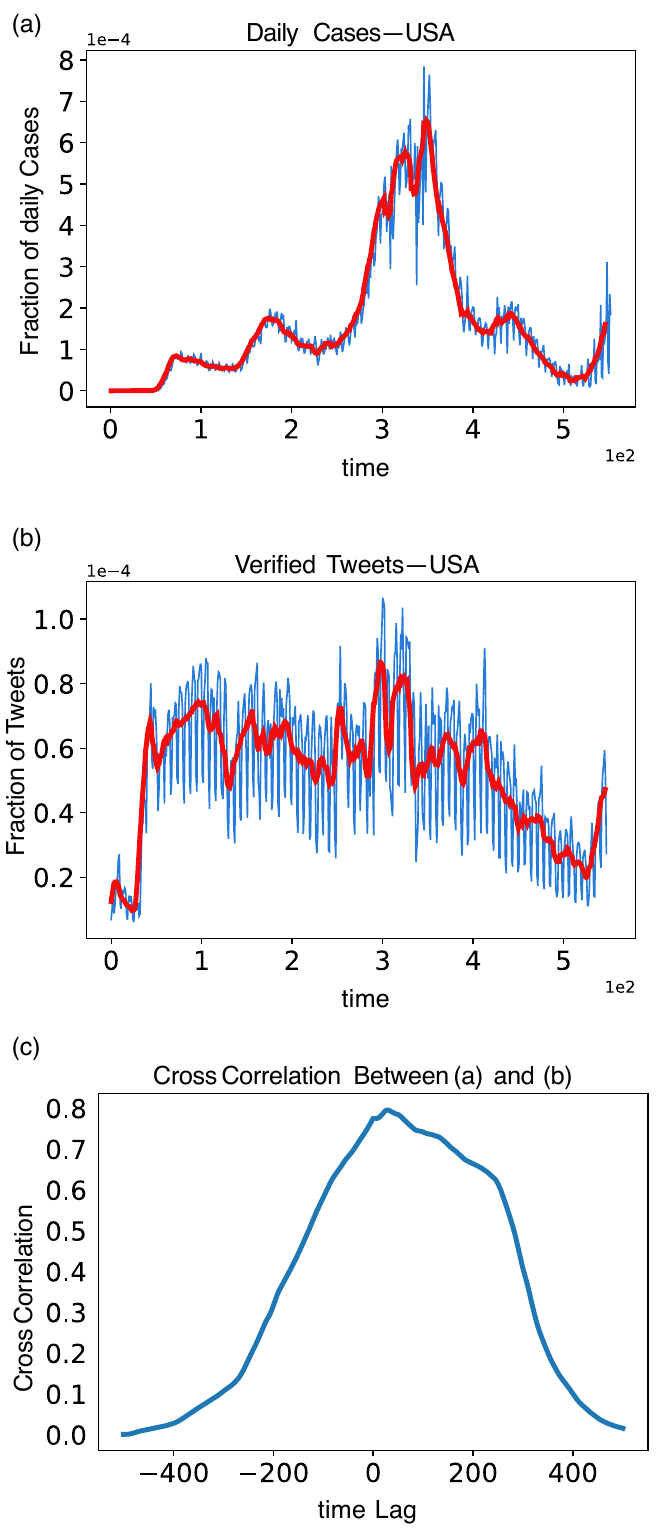}\\ 
    \caption{Epidemic and Infodemic Dynamics in the USA.
    (a) Daily reported COVID-19 cases in the USA. The blue line represents the raw daily case counts, while the orange line depicts a 7-day simple moving average for smoothing. Data sources are detailed in the Empirical Data section.
    (b) Daily count of verified tweets related to COVID-19 from users in the USA, processed using the same method as in panel (a).
    (c) Cross-correlation between the smoothed time series in panels (a) and (b), illustrating the relationship between epidemic spread and information dynamics.}
\label{fig:CaseCorr}
\end{figure}

\begin{figure*}
    \includegraphics[width=1.02\textwidth]{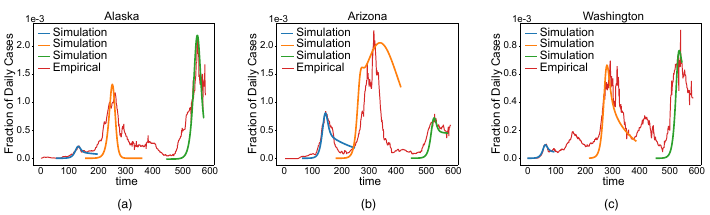}

    \caption{Fitting Our Model to Empirical Data for Three States: (a) Alaska, (b) Arizona, and (c) Washington, across three epidemic waves. Details for each peak of all states are provided in Table~\ref{tab:StateParameters} in Appendix~\ref{app:States}. }
    \label{fig:DeltaPeak}
\end{figure*}

\section{Empirical data and results}%
In this section, we introduce the sources of data collection, explain how we collected the data, and then proceed with the analysis of the graphs. We utilized two datasets for two different dynamics, which consisted of A. daily cases and B. data for prevalence, along with data from tweets for awareness. The temporal evolution of the fraction of daily cases and the fraction of tweets is illustrated in Fig.~\ref{fig:CaseCorr}(a) and Fig.~\ref{fig:CaseCorr}(b) respectively. Further details follow. 

\subsection{Daily cases:}
We obtained daily infection data from the information provided by the New York Times GitHub~\cite{Doe}, which includes cumulative infection reports segmented by country, state, and city on a daily basis, according to official sources. To enhance the analysis of the data, we applied a 7-day simple moving average method to smooth the entire dataset. 

\subsection{Twitter Activity (Awareness):}
Today, a large amount of people around the world use social networks. They share their ideas and life events on their pages. Although a very small percentage of information published in cyberspace is reliable~\cite{cuan2020misinformation}, by tracking and collecting this information and selecting the desired sections, the level of people's awareness can be attributed to them~\cite{funkmodelling}. For example, by using hashtag corona, information about coronavirus can be collected in a particular time. Of all the social networks used in the early days of the Coronavirus outbreak, Twitter was the most widely used in corona-related hashtags~\cite{cinelli2020covid}. Also twitter users share their posts in public pages more than other social media users and twitter gives more access to developers. The number of tweets can be represented as a state of awareness in society. 

Initially, we planned to collect awareness information from tweets by official news agencies and users on Twitter. However, this approach proved problematic due to two main issues: the presence of misinformation, which could distort public awareness and skew results, and the difficulty of geolocating tweets by city or state. To avoid misinformation and simplify data sourcing, we instead used state-level data from~\cite{gallotti2020assessing} to represent awareness more reliably. 

As discussed earlier, awareness can be modeled with a SIR dynamic. A recent study ~\cite{cinelli2020covid} showed fitting Twitter data with an SIR model gives appropriate results.

\begin{figure*}
    \centering    
    \includegraphics[width=1.02\linewidth]{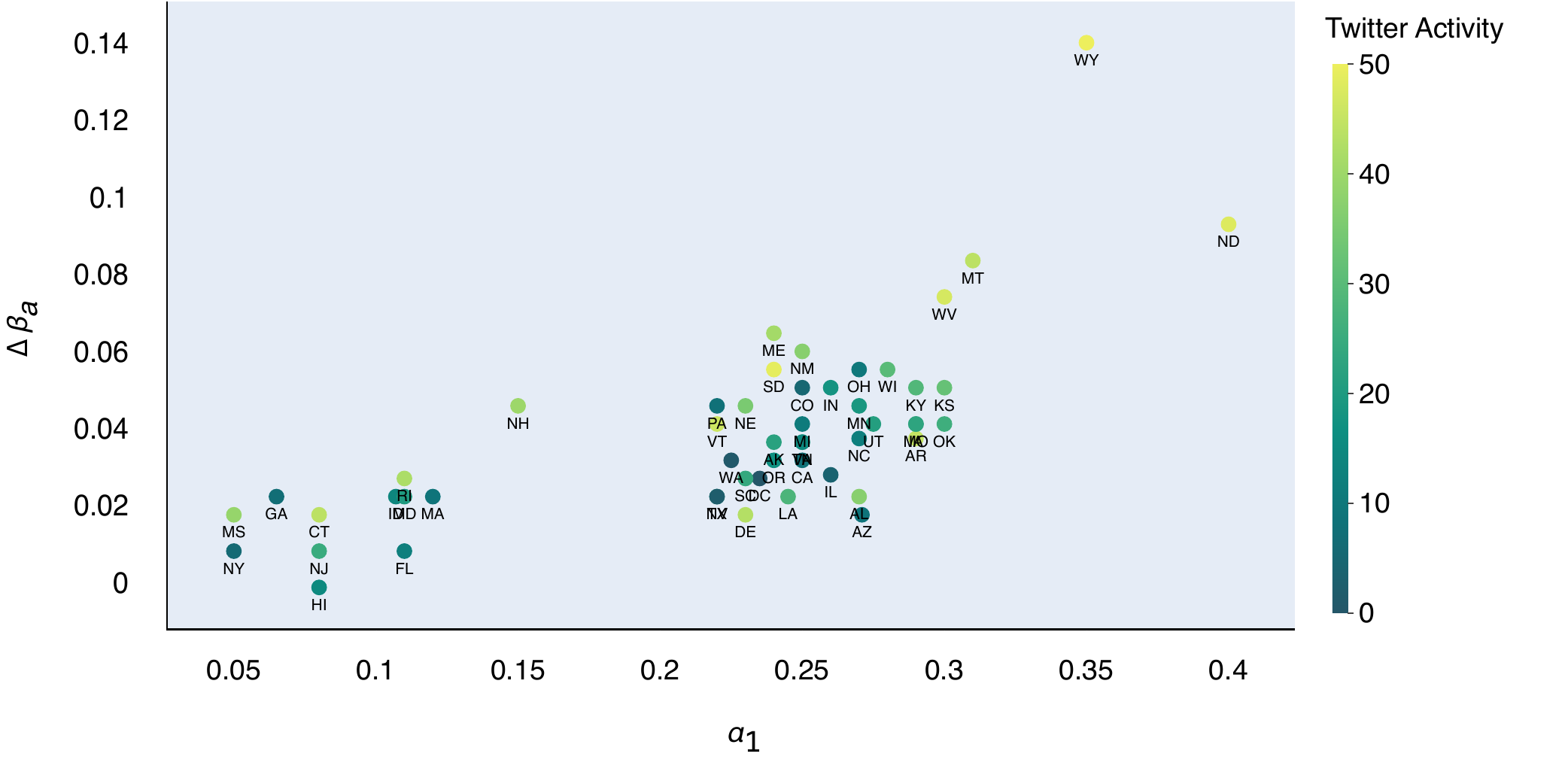}
    \caption{Correlation Between Four Variables Across U.S. States.
    The y-axis represents the change in the awareness transition rate between the first and second peaks, while the x-axis shows the immunity factor during the first peak. Both quantities were estimated by fitting the model to empirical data.
    The color scale indicates the Twitter activity ranking for each state. See also the correlations presented in Table~\ref{Tab:correlations}.}\label{fig:VarCorr}
\end{figure*}

\subsection{Empirical Results:}
We observe a strong correlation between the fraction of daily verified tweets and the fraction of daily new cases, see Fig.~\ref{fig:CaseCorr}(c). This strong correlation between awareness and disease dynamics is also evident in our simulation results for certain parameter values.
We can now fit the simulation results to empirical data for each state. This approach allows us to determine the values of the parameters $\beta_a$, $\alpha_1$, and $\alpha_2$.

We fitted our model to the first three waves of each state, with results shown for Alaska, Arizona, and Washington in Fig.\ref{fig:DeltaPeak}. The fitted parameters for all U.S. states can be found in Table\ref{tab:StateParameters}. We can see that the extracted parameters are in the region where we have a strong correlation; Also, the $\alpha_1$ for these states are near transition area for the epidemic.
Since we have different values for different US states, we can infer that the interplay of awareness and disease dynamics was different in the states. This result tells us that the spreading of information is not different in distinct society, but the impact of infodemic could be different.

In Fig ~\ref{fig:VarCorr}, we illustrate estimated parameters by US states for their first two peaks and their relation with the states' ranking by their twitter activity. 
It shows that the larger the immunity parameter $\alpha_{1}$ in the first peak, the greater the difference between the awareness transition rate in the first and second peaks, i.e. $\Delta\beta_{a}$. 
In other words, when immunity is stronger during the first peak, the awareness transition is more likely to be preserved in the dynamics up to the second peak. That is, a stronger influence of awareness on the disease can impact the rate of transition within the awareness dynamics itself. As shown in Fig.~\ref{fig:Equilibrium}(f), the awareness transition rate $\beta_{a}$ can be maintained at a fixed level by reducing $\alpha_{1}$. Table.~\ref{Tab:correlations} shows a strong correlation between $\alpha_{1}$ and $\Delta\beta_{a}$. This suggests that if awareness spreads at a higher rate, stronger immunity can be achieved by the time of the next peak. This is also confirmed by Fig.~\ref{fig:Equilibrium}(f), where moving toward a lower $\alpha_1$ and $\beta_a$ leads to higher immunity.

Additionally, examining the other correlations in Table~\ref{Tab:correlations} reveals the following insights:  
1) States that implemented mask orders earlier tend to exhibit stronger immunity at the first peak, and the awareness transition rate is more effectively preserved between the first two peaks.  
2) Twitter activity ranking appears to be related to the estimated parameters. Most states with higher Twitter activity (i.e., lower ranking values) tend to preserve the awareness transition rate during the first two peaks and also show higher immunity at the first peak.

\begin{table}[h!]
\centering

\begin{adjustbox}{width=0.5\textwidth}
\begin{tabular}{ |c|c|c| } 
\hline
Parameters/Indexes & Pearson Coef. & P values \\
\hline
$\Delta\beta_{a}$ and $\alpha_{1}$ & 0.688 & 2.410e-08 \\ 
$\Delta\alpha_{a}$ and $\beta_{a}$ & 0.575 & 9.981e-06 \\ 
$\Delta\beta_{a}$ and Twitter activity ranking & 0.476 & 4.039e-04\\
$\Delta\beta_{a}$ and Statewide mask orders & 0.522 & 8.256e-05 \\
$\Delta\alpha_{a}$ and Statewide mask orders & 0.470 & 4.939e-04 \\
\hline
\end{tabular}
\end{adjustbox}

\caption{Correlation Table, This table presents how the model parameters correlate across different epidemic waves, as well as with various state-level indices such as Twitter activity rankings and statewide mask mandate orders.}
\label{Tab:correlations}
\end{table}

\section{Discussion and Concluding Remarks}%
In this work, we introduced a coupled awareness–epidemic spreading model and analyzed its behavior using both simulated and empirical data. Our approach aimed to capture how information dynamics (infodemics) interact with epidemic processes in specific populations. Despite the simplicity of our assumptions—including the use of a mean-field approximation, population homogeneity, and a classical SIR framework—the model demonstrated meaningful alignment with real-world data in many cases. Nonetheless, some deviations were observed, underscoring the limitations of our simplified assumptions and highlighting areas for refinement.

We focused on the impact of local information—information exchanged through direct contacts or personal observations—on epidemic dynamics. While prior work has shown that global information, such as that disseminated via mass media, tends to reduce epidemic prevalence without affecting the epidemic threshold, our model is more concerned with threshold-shifting effects. For this reason, we excluded global information from the present analysis, although incorporating it in future work could yield insights into the early and transitional phases of disease dynamics before equilibrium is reached.

Our model assumes that awareness primarily influences the infection rate, rather than infectivity or recovery. This is based on the assumption that aware susceptible individuals are more likely to adopt preventive behaviors, while infected individuals may not significantly alter their behavior even when aware~\cite{funk2010endemic}. Other studies have explored broader roles for awareness, including its effect on recovery or immunity loss rates ~\cite{funkmodelling}.

While various pathways exist to model the effect of awareness—ranging from changes in network structure due to quarantine or vaccination~\cite{basu2008integrating, zanette2008infection} to assigning individual awareness levels with distinct dynamics~\cite{wu2012impact}—we chose a disease-like awareness propagation framework. This approach allows for a well-defined, parameterized model that aligns with empirical data and simplifies the analysis through a homogeneous mean-field approximation.

Despite these limitations, the framework is adaptable. Future extensions could incorporate heterogeneous mean-field models to better capture individual-level variations in awareness and disease susceptibility. Additionally, differentiating between local and global information mechanisms would allow for a more precise understanding of their respective impacts on epidemic control. Network-based simulations on realistic contact structures could further enhance the model's fidelity and allow for the study of behavioral interventions, structural changes, and vaccination strategies. 

Also this work aligns with and extends prior studies on interacting disease dynamics, such as ~\cite{khazaee2022effects, grassberger2016phase,ghanbarnejad2022emergence, chen2013outbreaks,rodriguez2017risk, pinotti2020interplay, sajjadi2021impact, zarei2019exact,rodriguez2019particle,cai2015avalanche,soriano2019markovian}. By focusing on awareness as a dynamic, disease-like process, we contribute to a growing body of literature that views behavioral responses as integral components of epidemic modeling.

In conclusion, our findings support the notion that awareness dynamics can significantly alter the course of an epidemic. In particular, our model enables the estimation of awareness-related parameters even from limited data sources, such as daily case numbers, making it a practical tool, see Fig.~\ref{fig:VarCorr}, for informing policy in data-constrained settings. Future works will enhance the model, further improving its relevance for epidemic control and public health decision-making.

\bibliographystyle{ieeetr}
\bibliography{citation}

\providecommand{\noopsort}[1]{}\providecommand{\singleletter}[1]{#1}%
\begin{thebibliography}{10}

\bibitem{ferguson2007capturing}
N.~Ferguson, ``Capturing human behaviour,'' {\em Nature}, vol.~446, no.~7137, pp.~733--733, 2007.

\bibitem{lv2021knowledge}
G.~Lv, J.~Yuan, S.~Hsieh, R.~Shao, and M.~Li, ``Knowledge and determinants of behavioral responses to the pandemic of covid-19,'' {\em Frontiers in Medicine}, vol.~8, 2021.

\bibitem{funk2010modelling}
S.~Funk, M.~Salath{\'e}, and V.~A. Jansen, ``Modelling the influence of human behaviour on the spread of infectious diseases: a review,'' {\em Journal of the Royal Society Interface}, vol.~7, no.~50, pp.~1247--1256, 2010.

\bibitem{saha2019modelling}
S.~Saha and G.~Samanta, ``Modelling and optimal control of hiv/aids prevention through prep and limited treatment,'' {\em Physica A: Statistical Mechanics and its Applications}, vol.~516, pp.~280--307, 2019.

\bibitem{cinelli2020covid}
M.~Cinelli, W.~Quattrociocchi, A.~Galeazzi, C.~M. Valensise, E.~Brugnoli, A.~L. Schmidt, P.~Zola, F.~Zollo, and A.~Scala, ``The covid-19 social media infodemic,'' {\em Scientific reports}, vol.~10, no.~1, pp.~1--10, 2020.

\bibitem{wu2018colored}
Q.~Wu and G.~Xiao, ``A colored mean-field model for analyzing the effects of awareness on epidemic spreading in multiplex networks,'' {\em Chaos: An Interdisciplinary Journal of Nonlinear Science}, vol.~28, no.~10, p.~103116, 2018.

\bibitem{funk2010endemic}
S.~Funk, E.~Gilad, and V.~A. Jansen, ``Endemic disease, awareness, and local behavioural response,'' {\em Journal of theoretical biology}, vol.~264, no.~2, pp.~501--509, 2010.

\bibitem{funk2009spread}
S.~Funk, E.~Gilad, C.~Watkins, and V.~A. Jansen, ``The spread of awareness and its impact on epidemic outbreaks,'' {\em Proceedings of the National Academy of Sciences}, vol.~106, no.~16, pp.~6872--6877, 2009.

\bibitem{wu2012impact}
Q.~Wu, X.~Fu, M.~Small, and X.-J. Xu, ``The impact of awareness on epidemic spreading in networks,'' {\em Chaos: an interdisciplinary journal of nonlinear science}, vol.~22, no.~1, p.~013101, 2012.

\bibitem{granell2014competing}
C.~Granell, S.~G{\'o}mez, and A.~Arenas, ``Competing spreading processes on multiplex networks: awareness and epidemics,'' {\em Physical review E}, vol.~90, no.~1, p.~012808, 2014.

\bibitem{kermack1927contribution}
W.~O. Kermack and A.~G. McKendrick, ``A contribution to the mathematical theory of epidemics,'' {\em Proceedings of the royal society of london. Series A, Containing papers of a mathematical and physical character}, vol.~115, no.~772, pp.~700--721, 1927.

\bibitem{bagnoli2007risk}
F.~Bagnoli, P.~Lio, and L.~Sguanci, ``Risk perception in epidemic modeling,'' {\em Physical Review E}, vol.~76, no.~6, p.~061904, 2007.

\bibitem{linka2020reproduction}
K.~Linka, M.~Peirlinck, and E.~Kuhl, ``The reproduction number of covid-19 and its correlation with public health interventions,'' {\em Computational mechanics}, vol.~66, pp.~1035--1050, 2020.

\bibitem{Doe}
https://github.com/nytimes/covid 19-data, ``Coronavirus (covid-19) data in the united states.''

\bibitem{cuan2020misinformation}
J.~Y. Cuan-Baltazar, M.~J. Mu{\~n}oz-Perez, C.~Robledo-Vega, M.~F. P{\'e}rez-Zepeda, and E.~Soto-Vega, ``Misinformation of covid-19 on the internet: infodemiology study,'' {\em JMIR public health and surveillance}, vol.~6, no.~2, p.~e18444, 2020.

\bibitem{funkmodelling}
S.~Funk, M.~Salath{\'e}, and V.~A. Jansen, ``Modelling the influence of human behaviour on the spread of,''

\bibitem{gallotti2020assessing}
R.~Gallotti, F.~Valle, N.~Castaldo, P.~Sacco, and M.~De~Domenico, ``Assessing the risks of ‘infodemics’ in response to covid-19 epidemics,'' {\em Nature Human Behaviour}, vol.~4, no.~12, pp.~1285--1293, 2020.

\bibitem{basu2008integrating}
S.~Basu, G.~B. Chapman, and A.~P. Galvani, ``Integrating epidemiology, psychology, and economics to achieve hpv vaccination targets,'' {\em Proceedings of the national academy of sciences}, vol.~105, no.~48, pp.~19018--19023, 2008.

\bibitem{zanette2008infection}
D.~H. Zanette and S.~Risau-Gusm{\'a}n, ``Infection spreading in a population with evolving contacts,'' {\em Journal of biological physics}, vol.~34, no.~1, pp.~135--148, 2008.

\bibitem{khazaee2022effects}
A.~Khazaee and F.~Ghanbarnejad, ``Effects of measures on phase transitions in two cooperative susceptible-infectious-recovered dynamics,'' {\em Physical Review E}, vol.~105, no.~3, p.~034311, 2022.

\bibitem{grassberger2016phase}
P.~Grassberger, L.~Chen, F.~Ghanbarnejad, and W.~Cai, ``Phase transitions in cooperative coinfections: Simulation results for networks and lattices,'' {\em Physical Review E}, vol.~93, no.~4, p.~042316, 2016.

\bibitem{ghanbarnejad2022emergence}
F.~Ghanbarnejad, K.~Seegers, A.~Cardillo, and P.~H{\"o}vel, ``Emergence of synergistic and competitive pathogens in a coevolutionary spreading model,'' {\em Physical Review E}, vol.~105, no.~3, p.~034308, 2022.

\bibitem{chen2013outbreaks}
L.~Chen, F.~Ghanbarnejad, W.~Cai, and P.~Grassberger, ``Outbreaks of coinfections: The critical role of cooperativity,'' {\em EPL (Europhysics Letters)}, vol.~104, no.~5, p.~50001, 2013.

\bibitem{rodriguez2017risk}
J.~P. Rodr{\'\i}guez, F.~Ghanbarnejad, and V.~M. Egu{\'\i}luz, ``Risk of coinfection outbreaks in temporal networks: A case study of a hospital contact network,'' {\em Frontiers in Physics}, vol.~5, p.~46, 2017.

\bibitem{pinotti2020interplay}
F.~Pinotti, F.~Ghanbarnejad, P.~H{\"o}vel, and C.~Poletto, ``Interplay between competitive and cooperative interactions in a three-player pathogen system,'' {\em Royal Society open science}, vol.~7, no.~1, p.~190305, 2020.

\bibitem{sajjadi2021impact}
S.~Sajjadi, M.~R. Ejtehadi, and F.~Ghanbarnejad, ``Impact of temporal correlations on high risk outbreaks of independent and cooperative sir dynamics,'' {\em Plos one}, vol.~16, no.~7, p.~e0253563, 2021.

\bibitem{zarei2019exact}
F.~Zarei, S.~Moghimi-Araghi, and F.~Ghanbarnejad, ``Exact solution of generalized cooperative susceptible-infected-removed (sir) dynamics,'' {\em Physical Review E}, vol.~100, no.~1, p.~012307, 2019.

\bibitem{rodriguez2019particle}
J.~P. Rodr{\'\i}guez, F.~Ghanbarnejad, and V.~M. Egu{\'\i}luz, ``Particle velocity controls phase transitions in contagion dynamics,'' {\em Scientific reports}, vol.~9, no.~1, pp.~1--9, 2019.

\bibitem{cai2015avalanche}
W.~Cai, L.~Chen, F.~Ghanbarnejad, and P.~Grassberger, ``Avalanche outbreaks emerging in cooperative contagions,'' {\em Nature physics}, vol.~11, no.~11, pp.~936--940, 2015.

\bibitem{soriano2019markovian}
D.~Soriano-Pa{\~n}os, F.~Ghanbarnejad, S.~Meloni, and J.~G{\'o}mez-Garde{\~n}es, ``Markovian approach to tackle the interaction of simultaneous diseases,'' {\em Physical Review E}, vol.~100, no.~6, p.~062308, 2019.

\bibitem{hethcote2002mathematics}
H.~W. Hethcote, ``The mathematics of infectious diseases,'' {\em Mathematical biosciences}, vol.~180, no.~1-2, pp.~1--21, 2002.

\end{thebibliography}

\newpage
\newpage
\onecolumngrid
\appendix

\clearpage

\section{Basic Reproductive Ratio and Effective Reproductive Ratio}
\label{app:R_0}
The basic reproductive ratio, $R_0$, is defined as the expected number of secondary cases generated by a typical infectious individual in a completely susceptible population~\cite{hethcote2002mathematics}. It plays a crucial role in characterizing the initial spread and prevalence of an infectious disease.

However, in our model, a more informative metric is the effective reproduction number, $R_t$, which captures time-varying transmission dynamics as the susceptible population becomes increasingly aware and, therefore, less likely to contribute to transmission. We employ the next generation matrix method to compute both $R_0$ and $R_t$.

\begin{center}
\[
    F = 
    \begin{bmatrix}
    \beta_i [US_0] & \beta_i [US_0] & \beta_i [US_0] \\
    \hat{\beta}_i [A_1S_0] & \hat{\beta}_i [A_1S_0] & \hat{\beta}_i [A_1S_0] \\
    \hat{\beta}_i [A_2S_0] & \hat{\beta}_i [A_2S_0] & \hat{\beta}_i [A_2S_0]
    \end{bmatrix}
\]
\end{center}

\begin{center}
\[
    V =
    \begin{bmatrix}
    \hat{\beta}_a [A_1S_0] + \gamma_i & 0 & 0 \\
    - \hat{\beta}_a [A_1S_0] & \gamma_a + \gamma_i & 0 \\
    0 & -\gamma_a & \gamma_i
    \end{bmatrix}
\]
\end{center}

\begin{center}
\small
\begin{align*}
FV^{-1} = \frac{\beta_{i}}{\gamma_{i}}
\begin{bmatrix}
\beta_i [US_0] \left( \dfrac{1}{\hat{\beta}_a [A_1S_0] + \gamma_i} \right)
& \beta_i [US_0] \left( \dfrac{\hat{\beta}_a [A_1S_0]}{(\hat{\beta}_a [A_1S_0] + \gamma_i)(\gamma_a + \gamma_i)} \right)
& \beta_i [US_0] \left( \dfrac{\hat{\beta}_a [A_1S_0] \gamma_a}{(\hat{\beta}_a [A_1S_0] + \gamma_i)(\gamma_a + \gamma_i)\gamma_i} \right)
\\[1.5ex]
\hat{\beta}_i [A_1S_0] \left( \dfrac{1}{\hat{\beta}_a [A_1S_0] + \gamma_i} \right)
& \hat{\beta}_i [A_1S_0] \left( \dfrac{\hat{\beta}_a [A_1S_0]}{(\hat{\beta}_a [A_1S_0] + \gamma_i)(\gamma_a + \gamma_i)} + \dfrac{1}{\gamma_a + \gamma_i} \right)
& \hat{\beta}_i [A_1S_0] \left( \dfrac{\hat{\beta}_a [A_1S_0] \gamma_a}{(\hat{\beta}_a [A_1S_0] + \gamma_i)(\gamma_a + \gamma_i)\gamma_i} + \dfrac{\gamma_a}{(\gamma_a + \gamma_i)\gamma_i} \right)
\\[1.5ex]
\hat{\beta}_i [A_2S_0] \left( \dfrac{1}{\hat{\beta}_a [A_1S_0] + \gamma_i} \right)
& \hat{\beta}_i [A_2S_0] \left( \dfrac{\hat{\beta}_a [A_1S_0]}{(\hat{\beta}_a [A_1S_0] + \gamma_i)(\gamma_a + \gamma_i)} + \dfrac{1}{\gamma_a + \gamma_i} \right)
& \hat{\beta}_i [A_2S_0] \left( \dfrac{\hat{\beta}_a [A_1S_0] \gamma_a}{(\hat{\beta}_a [A_1S_0] + \gamma_i)(\gamma_a + \gamma_i)\gamma_i} + \dfrac{\gamma_a}{(\gamma_a + \gamma_i)\gamma_i} + \dfrac{1}{\gamma_i} \right)
\end{bmatrix}
\end{align*}
\normalsize
\end{center}

\begin{center}
\begin{align*}
R_0 =\ & \beta_i [US_0] \left( \frac{1}{\hat{\beta}_a [A_1S_0] + \gamma_i}
+ \frac{\hat{\beta}_a [A_1S_0]}{(\hat{\beta}_a [A_1S_0] + \gamma_i)(\gamma_a + \gamma_i)}
+ \frac{\hat{\beta}_a [A_1S_0] \gamma_a}{(\hat{\beta}_a [A_1S_0] + \gamma_i)(\gamma_a + \gamma_i)\gamma_i} \right) \\
& + \hat{\beta}_i [A_1S_0] \left( \frac{1}{\gamma_a + \gamma_i} + \frac{\gamma_a}{(\gamma_a + \gamma_i)\gamma_i} \right)
+ \hat{\beta}_i [A_2S_0] \cdot \frac{1}{\gamma_i}
\end{align*}

\end{center}

\begin{center}
\begin{align*}
    R_0^{\text{eff}} = \lambda_{\max} = \frac{\beta_{i}}{\gamma_{i}} \left( [US_0] + \alpha_{1} [A_{1}S_0] \right)
\end{align*}
\end{center}

This expression for $R_0^{\text{eff}}$ depends on the immunization level $\alpha_1$. In the case of no immunization ($\alpha_1 = 1$), the model reduces to the standard SIR form with a fully susceptible population. Conversely, when full immunization is achieved among the aware individuals ($\alpha_1 = 0$), only the unaware susceptible group $[US_0]$ contributes to the effective reproduction number.

The corresponding time-dependent effective reproduction number is given by:

\begin{align*}
    R_t^{\text{eff}} = \frac{\beta_{i}}{\gamma_{i}} \left( [US] + \alpha_{1} \left( [A_{1}S] + [A_{2}S] \right) \right)
\end{align*}

This formulation illustrates how increasing awareness and immunization reduce transmission potential over time. Notably, $R_t^{\text{eff}}$ does not explicitly depend on the immunization parameter $\alpha_2$, as demonstrated by the simulation results in Fig.~\ref{fig:Rstar}(a). 

Additionally, $R_t^{\text{eff}}$ does not explicitly depend on the awareness transmission rate $\beta_a$. However, varying $\beta_a$ alters the distribution of the susceptible population by replacing unaware susceptibles $[US]$ with aware susceptibles ($[A_1S]$ and $[A_2S]$). If $\alpha_1$ is not equal to 1 (i.e., there is some immunization), this shift results in a decreased value of $R_t^{\text{eff}}$ over time, as shown in Fig.~\ref{fig:Rstar}(b) and Fig.~\ref{fig:Rstar}(c).

\newpage
\section{States}
\label{app:States}
 
\begin{table*}[h]
  \centering
  \caption{The parameters were extracted by fitting the model to empirical COVID-19 data for each U.S. state, with associated errors reflecting the uncertainty in the fitting process.}
  \label{tab:StateParameters}
    \begin{tabular}{ |p{1.6cm}|p{0.7cm}|p{2.1cm}|p{2.1cm}|p{2.1cm}|p{2.1cm}|     p{2.1cm}| p{2.1cm}|p{1cm}|}
    
    \hline
    State Name & abb &
     $\alpha_1$ p1 & $\beta_a$ p1 & $\alpha_1$ p2 & $\beta_a$ p2 &    $\alpha_1$ p3 & $\beta_a$ p3 & ranking 
     \\
     \hline
     Alaska	& AK & 
     0.240 (0.2453,0.2490) & 
     0.215 (0.2145,0.2152) &
     0.160 (0.1686,0.1734) &
     0.185 (0.185,0.185) &
     0.021 (0.0146,0.0240) &
     0.175 (0.1749,0.1750) &
     22

    \\

     Alabama & AL & 
     0.270 (0.2704,0.2756) &
     0.210 (0.2095,0.2104) &
     0.254 (0.2522,0.2558) &
     0.195 (0.1950,0.1950) &
     0.139 (0.1361,0.1546) &
     0.190 (0.1895, 0.1901) &
     37 
     
    \\
    
    Arkansas & AR &
    0.290 (0.2883,0.2950) &
    0.226 (0.2239,0.2268) &
    0.182 (0.1810,0.1863) &
    0.195 (0.1950,0.1950) &
    0.167 (0.1649,0.1777) &
    0.200 (0.1999,0.2008) &
    45 

    \\
    Arizona	& AZ &
    0.271 (0.2703,0.2750) &
    0.205 (0.2045,0.2052) &
    0.330 (0.3297,0.3310) &
    0.195 (0.1950,0.1950) &
    0.199 (0.1998,0.2055) &
    0.200 (0.2000,0.2000) &
    9 
     
    \\
    California & CA &
 
    0.250 (0.2486,0.2547) &
    0.240 (0.2337,0.2356) &
    0.217 (0.2158,0.2209) &
    0.215 (0.2150,0.2150) &
    0.239 (0.2377,0.2476) &
    0.235 (0.2348,0.2368) &
    8 
    
    \\
    
    Colorado & CO &
    0.250 (0.2485,0.2542) &
    0.235 (0.2334,0.2353) &
    0.135 (0.1329,0.1517) &
    0.190 (0.1895,0.1902) &
    0.145 (0.1429,0.1604) &
    0.205 (0.2045,0.2052) &
    5 
    
    \\

    Connecticut & CT &
    0.080 (0.0772,0.0968) &
    0.205 (0.2048,0.2055) &
    0.245 (0.2435,0.2472) &
    0.195 (0.1950,0.1950) &
    0.196 (0.1931,0.2076) &
    0.205 (0.2042,0.2051) &
    44 
    
    \\

    Delaware & DE &
    0.235 (0.2325,0.2441) &
    0.215 (0.2137,0.2150) &
    0.237 (0.2356,0.2397) &
    0.195 (0.1950,0.1950) &
    0.149 (0.1488,0.1958) &
    0.200 (0.1997,0.2016) &
    23 
    
    \\

    District of Columbia & DC & 
    0.230 (0.2210,0.2376) &
    0.215 (0.2123,0.2141) &
    0.227 (0.2254,0.2312) &
    0.205 (0.2045,0.2052) &
    0.155 (0.1511,0.1702) &
    0.205 (0.2045,0.2055) &
    43 
    
    \\
    
    Florida & FL &
    0.019 (0.1077,0.1176) &
    0.195 (0.1950,0.1950) &
    0.236 (0.2342,0.2378) &
    0.195 (0.1950,0.1950) &
    0.148 (0.1476,0.1537) &
    0.185 (0.1850,0.1850) &
    12 
    
    \\

    Georgia	& GA &
    0.065 (0.0567,0.0953) &
    0.205 (0.2051,0.2065) &
    0.207 (0.2050,0.2109) &
    0.190 (0.1900,0.1900) &
    0.150 (0.1444,0.1982) &
    0.190 (0.1902,0.1923) &
    7 
    
    \\

    Hawaii & HI &
    0.080 (0.0726,0.1228) &
    0.215 (0.2125,0.2148) &
    0.090 (0.0892,0.1328) &
    0.225 (0.2233,0.2257) &
    0.010 (0.0890,0.1370) &
    0.195 (0.1958,0.1975) &
    15 
    
    \\

    Iowa & IA & 
    0.290 (0.2790,0.2830) &
    0.220 (0.2195,0.2211) &
    0.198 (0.1960,0.2013) &
    0.185 (0.1850,0.1850) &
    0.080 (0.0683,0.1097) &
    0.195 (0.1951,0.1965) &
    23 
    
    \\

    Idaho & ID &
    0.107 (0.1051,0.1456) & 
    0.205 (0.2042,0.2058) & 
    0.208 (0.2067,0.2112) & 
    0.190 (0.1900,0.1900) & 
    0.100 (0.0836,0.1451) & 
    0.192 (0.1917,0.1936) &
    14 
    
    \\
    
    West Virginia & WV &
    0.300 (0.2973,0.3074) &
    0.260 (0.2534,0.2602) &
    0.188 (0.1868,0.1925) &
    0.190 (0.1900,0.1900) &
    0.148 (0.1479,0.1548) &
    0.185 (0.1850,0.1850) &
    47 
    
    \\
    
    Wisconsin & WI & 
    0.280 (0.2775,0.2845) &
    0.235 (0.2315,0.2338) &
    0.195 (0.1949,0.1997) &
    0.185 (0.1850,0.1850) &
    0.320 (0.3055,0.3498) &
    0.205 (0.2041,0.2066) &
    30 
    
    \\
    
    Washington & WA &
    0.225 (0.2260,0.2388) &
    0.230 (0.2265,0.2288) &
    0.215 (0.2116,0.2243) &
    0.205 (0.2042,0.2051) &
    0.180 (0.1639,0.2201) &
    0.200 (0.1993,0.2014) &
    1 
    
    \\
    
    Virginia & VA & 
    0.250 (0.2484,0.2536) &
    0.225 (0.2248,0.2268) &
    0.245 (0.2442,0.2478) &
    0.195 (0.1950,0.1950) &
    0.150 (0.1077,0.1883) &
    0.200 (0.2016,0.2044) &
    25 
     
    \\
    
    Utah & UT & 
    0.275 (0.2721,0.2792) &
    0.225 (0.2247,0.2250) &
    0.235 (0.2329,0.2371) &
    0.190 (0.1900,0.1900) &
    0.305 (0.2571,0.4002) &
    0.205 (0.2010,0.2070) &
    21 
    
    \\
    
    Illinois & IL &

    0.260 (0.2573,0.2674) &
    0.216 (0.2176,0.2193) &
    0.300 (0.2999,0.3021) &
    0.195 (0.1950,0.1950) &
    0.180 (0.1773,0.1854) &
    0.205 (0.2050,0.2050) &
    4

    \\
    
    Indiana & IN &

    0.260 (0.2568,0.2645) &
    0.230 (0.2301,0.2319) &
    0.085 (0.0826,0.0894) &
    0.185 (0.1850,0.1850) &
    0.030 (0.0228,0.0485) &
    0.190 (0.1899,0.1908) &
    18 
    
    \\
    
    Kansas & KS &
    0.300 (0.2911,0.3002) &
    0.235 (0.2334,0.2382) &
    0.180 (0.1786,0.1847) &
    0.190 (0.1900,0.1900) &
    0.120 (0.1167,0.1340) &
    0.195 (0.1945,0.1952) &
    32 
    
    \\
    
    Kentucky & KY &
    0.290 (0.2812,0.2908) &
    0.240 (0.2397,0.2440) &
    0.245 (0.2431,0.2469) &
    0.195 (0.1950,0.1950) &
    0.040 (0.0373,0.0487) &
    0.185 (0.1850,0.1850) &
    29
    
    \\
    
    Louisiana & LA &
    0.245 (0.2450,0.2509) &
    0.210 (0.2095,0.2102) &
    0.225 (0.2256,0.2297) &
    0.195 (0.1950,0.1950) &
    0.190 (0.1881,0.2433) &
    0.185 (0.1856,0.1873) &
    28
    
    \\
    
    Massachusetts & MA &
    0.120 (0.1115,0.1378) &
    0.205 (0.2044,0.2059) &
    0.252 (0.2516,0.2551) &
    0.190 (0.1900,0.1900) &
    0.220 (0.2173,0.2440) &
    0.210 (0.2093,0.2111) &
    9 
    
    \\
    
    Maryland & MD & 
    0.112 (0.1105,0.1421) &
    0.215 (0.2133,0.2150) &
    0.227 (0.2263,0.2310) &
    0.200 (0.2000,0.2000) &
    0.100 (0.0749,0.1144) &
    0.210 (0.2096,0.2117) &
    20

    \\
    
    \hline

    \end{tabular}

\end{table*}

\newpage

\begin{table*}[h]
  \centering
  \caption*{}
  \label{tab2}
    \begin{tabular}{ |p{1.5cm}||p{0.5cm}|p{2.1cm}|p{2.1cm}|p{2.1cm}|p{2.1cm}|     p{2.1cm}| p{2.1cm}|p{1cm}|}
    
    \hline

    Maine &	ME &
    0.248 (0.2468,0.2525) &
    0.253 (0.2513,0.2543) &
    0.247 (0.2448,0.2499) &
    0.205 (0.2045,0.2052) &
    0.150 (0.1314,0.1786) &
    0.200 (0.1997,0.2016) &
    41 

    \\

    Michigan & MI &
    0.250 (0.2471,0.2529) &
    0.221 (0.2206,0.2223) &
    0.030 (0.0204,0.0356) &
    0.190 (0.1900,0.1900) &
    0.120 (0.1187,0.1266) &
    0.190 (0.1900,0.1900) &
    11 
    
    \\
   
    Minnesota & MN &
    0.270 (0.2643,0.2704) &
    0.230 (0.2294,0.2316) &
    0.125 (0.1233,0.1327) &
    0.185 (0.1850,0.1850) &
    0.100 (0.0850,0.1323) &
    0.195 (0.1955,0.1971) &
    19 
    
    \\ 

    Missouri &	MO &
    0.290 (0.2871,0.2949) &
    0.230 (0.2279,0.2314) &
    0.280 (0.2797,0.2810) &
    0.195 (0.1950,0.1950) &
    0.225 (0.2280,0.2307) &
    0.205 (0.2050,0.2050) &
    23 
    
    \\
   
    Mississippi & MS &
    0.050 (0.0351,0.0663) &
    0.200 (0.1999,0.2011) &
    0.168 (0.1668,0.1739) &
    0.190 (0.1900,0.1900) &
    0.200 (0.1983,0.2816) &
    0.188 (0.1872,0.1901) &
    39 
    
    \\
    
    Montana & MT &
    0.310 (0.3098,0.3222) &
    0.265 (0.2614,0.2705) &
    0.090 (0.0899,0.0980) &
    0.185 (0.1850,0.1850) &
    0.140 (0.1088,0.1772) &
    0.187 (0.1864,0.1883) &
    44 
    
    \\
    
    North Carolina & NC & 
    0.270 (0.2653,0.2714) &
    0.226 (0.2256,0.2273) &
    0.235 (0.2345,0.2381) &
    0.195 (0.1950,0.1950) &
    0.100 (0.0663,0.1270) &
    0.190 (0.1909,0.1927) &
    12 
    
    \\
    
    North Dakota & ND &
    0.400 (0.3860,0.4006) &
    0.285 (0.2722,0.2868) &
    0.360 (0.3600,0.3600) &
    0.195 (0.1950,0.1950) &
    0.080 (0.0556,0.1111) &
    0.190 (0.1908,0.1925) &
    48 
    
    \\
    
    Nebraska & NE & 
    0.233 (0.2338,0.2415) &
    0.219 (0.2178,0.2194) &
    0.115 (0.1085,0.1169) &
    0.180 (0.1800,0.1800) &
    0.150 (0.1460,0.1666) &
    0.195 (0.1945,0.1952) &
    35 
    
    \\
    
    New Hampshire & NH &
    0.150 (0.1508,0.1885) &
    0.235 (0.2347,0.2380) &
    0.250 (0.2498,0.2568) &
    0.200 (0.1995,0.2002) &
    0.080 (0.0563,0.1023) &
    0.200 (0.2005,0.2021) &
    40 
    
    \\
    
    New Jersey & NJ &
    0.080 (0.0713,0.0920) & 
    0.200 (0.1995,0.2005) & 
    0.275 (0.2732,0.2774) & 
    0.200 (0.2000,0.2000) & 
    0.150 (0.1426,0.1834) & 
    0.210 (0.2094,0.2113) & 
    25
    
    \\
    
    New Mexico & NM & 
    0.250 (0.2444,0.2529) & 
    0.230 (0.2290,0.2313) & 
    0.020 (0.0120,0.0226) & 
    0.180 (0.1800,0.1800) & 
    0.240 (0.2212,0.2674) & 
    0.205 (0.2038,0.2062) & 
    37 
    
    \\
    
    Nevada & NV &
    0.220 (0.2202,0.2238) &
    0.205 (0.2050,0.2050) &
    0.200 (0.1970,0.2017) &
    0.190 (0.1900,0.1900) &
    0.225 (0.2243,0.2336) &
    0.205 (0.2045,0.2052) &
    22 
    
    \\
    
    New York & NY & 
    0.050 (0.0326,0.0621) & 
    0.195 (0.1949,0.1961) & 
    0.252 (0.2510,0.2543) & 
    0.195 (0.1950,0.1950) & 
    0.100 (0.0946,0.1454) & 
    0.210 (0.2091,0.2116) & 
    6 
    
    \\
    
    Ohio & OH & 
    0.270 (0.2659,0.2728) & 
    0.235 (0.2334,0.2362) & 
    0.040 (0.0403,0.0503) & 
    0.185 (0.1850,0.1850) & 
    0.080 (0.0638,0.1082) & 
    0.195 (0.1952,0.1967) & 
    10 
    
    \\
    
    Oklahoma & OK &
    0.290 (0.2900,0.2946) &
    0.220 (0.2200,0.2217) &
    0.216 (0.2157,0.2190) &
    0.190 (0.1900,0.1900) &
    0.115 (0.1141,0.1505) &
    0.195 (0.1950,0.1963) &
    26 
    
    \\
    
    Oregon & OR & 
    0.245 (0.2422,0.2472) &
    0.235 (0.2336,0.2354) &
    0.210 (0.2091,0.2182) &
    0.205 (0.2045,0.2052) &
    0.150 (0.1375,0.1805) &
    0.195 (0.1954,0.1969) &
    17
    \\
    Pennsylvania & PA & 
    0.230 (0.2304,0.2369) & 
    0.225 (0.2271,0.2289) & 
    0.150 (0.1470,0.1556) & 
    0.190 (0.1900,0.1900) & 
    0.150 (0.1463,0.1856) & 
    0.200 (0.2001,0.2019) & 
    8 
    
    \\
    
    Rhode Island & RI &
    0.110 (0.1097,0.1363) &
    0.205 (0.2043,0.2057) &
    0.175 (0.1750,0.1809) &
    0.185 (0.1850,0.1850) &
    0.130 (0.1239,0.1541) &
    0.205 (0.2035,0.2051) &
    42 
    
    \\
    
    South Carolina & SC &
    0.240 (0.2388,0.2418) & 
    0.210 (0.2100,0.2100) & 
    0.225 (0.2257,0.2290) & 
    0.190 (0.1900,0.1900) & 
    0.230 (0.2015,0.2885) & 
    0.188 (0.1871,0.1899) & 
    24 
    
    \\
    
    South Dakota & SD &
    0.245 (0.2429,0.2491) &
    0.226 (0.2266,0.2284) &
    0.150 (0.1473,0.1546) &
    0.180 (0.1800,0.1800) &
    0.050 (0.0297,0.0736) &
    0.195 (0.1949,0.1965) &
    49 
    
    \\
    
    Tennessee & TN &
    0.250 (0.2512,0.2581) &
    0.215 (0.2135,0.2149) &
    0.160 (0.1567,0.1640) &
    0.185 (0.1850,0.1850) &
    0.040 (0.0334,0.0466) &
    0.185 (0.1850,0.1850) &
    15 
    
    \\
    
    Texas & TX &
    0.225 (0.2244,0.2329) &
    0.210 (0.2092,0.2101) &
    0.245 (0.2425,0.2462) &
    0.195 (0.1950,0.1950) &
    0.140 (0.1360,0.1913) &
    0.195 (0.1948,0.1968) &
    2 
    
    \\
    
    Vermont	& VT &
    0.230 (0.2294,0.2386) &
    0.250 (0.2510,0.2540) &
    0.250 (0.2485,0.2542) &
    0.215 (0.2130,0.2146) &
    0.120 (0.1117,0.1470) &
    0.200 (0.1997,0.2010) &
    46 
    
    \\
    
    Wyoming	& WY &
    0.355 (0.3550,0.3803) &
    0.320 (0.2939,0.3270) &
    0.120 (0.1164,0.1236) &
    0.180 (0.1800,0.1800) &
    0.020 (0.0132,0.0341) &
    0.185 (0.1848,0.1855) &
    50 
    \\
    \hline
 
    \end{tabular}

\end{table*}

\vspace{1.0cm}
\end{document}